\RequirePackage{xcolor}

\documentclass[journal,twoside,web]{IEEEtran} 
\usepackage{cite}
\usepackage{amsmath,amssymb,amsfonts}
\usepackage{algorithmic}
\usepackage{graphicx}
\usepackage{textcomp}
\def\BibTeX{{\rm B\kern-.05em{\sc i\kern-.025em b}\kern-.08em
    T\kern-.1667em\lower.7ex\hbox{E}\kern-.125emX}}
\usepackage{acronym}

\usepackage[uncertainty-mode = separate,separate-uncertainty-units = single, mode=match]{siunitx}
\usepackage{cleveref}
\crefname{equation}{}{}
\Crefname{equation}{Equation}{Equations}
\crefname{table}{Table}{Table}
\crefname{figure}{Fig.}{Fig.}
\usepackage{tikz}
\usetikzlibrary{arrows.meta, decorations.pathmorphing, decorations.markings, shapes, shadings,arrows}
\usepackage{physics}
\usepackage{tabularx}
\usepackage{booktabs}
\usepackage{balance}
\usepackage{dsfont} 
\usepackage{subcaption}
\newcommand{\prob}{\mathbb{P}}
\newcommand{\point}[1]{\mathrm{#1}}
\newcommand{\diff}[1]{\overline{#1}}

\begin{document}
\bstctlcite{IEEEexample:BSTcontrol}

\title{Analytical Model of Prompt Gamma Timing for Spatiotemporal Emission Reconstruction in Particle Therapy}

\author{Julius~Werner,~\IEEEmembership{Graduate Student Member, IEEE}, Malte~Schmidt, Francesco~Pennazio, Jorge~Roser,~\IEEEmembership{Member, IEEE}, Jona Kasprzak,~\IEEEmembership{Graduate Student Member, IEEE}, Veronica~Ferrero, and Magdalena~Rafecas,~\IEEEmembership{Senior Member, IEEE}
\thanks{Manuscript received 28 October 2025. This work was supported in part by the German Research Foundation (DFG) by Grants No. 516587313 (PROSIT) and No. 383681334 (COMMA). (Corresponding author: Julius Werner).}
\thanks{This work did not involve human subjects or animals in its research.}
\thanks{Julius Werner, Malte Schmidt, Jorge Roser, Jona Kasprzak, and Magdalena Rafecas are with the Institute of Medical Engineering, Universität zu Lübeck, Ratzeburger Allee 160, 23562 Lübeck, Germany (e-mail: friedemann.werner@uni-luebeck.de; m.schmidt@uni-luebeck.de; jorge.rosermartinez@uni-luebeck.de; j.kasprzak@uni-luebeck.de; magdalena.rafecas@uni-luebeck.de).}
\thanks{Francesco Pennazio is with the Istituto Nazionale di Fisica Nucleare, Sezione di Torino, Via P. Giuria 1, 10125 Torino, Italy. (e-mail: francesco.pennazio@to.infn.it).}
\thanks{Veronica Ferrero is with the Department of Physics, Università degli Studi di Torino, Via P. Giuria 1, 10125 Torino, Italy, and the Istituto Nazionale di Fisica Nucleare, Sezione di Torino, Via P. Giuria 1, 10125 Torino, Italy (e-mail: veronica.ferrero@unito.it).}
}

\maketitle

\begin{abstract}
Particle therapy relies on up-to-date knowledge of the stopping power of the patient tissues to deliver the prescribed dose distribution. The stopping power describes the average particle motion, which is encoded in the distribution of prompt-gamma photon emissions in time and space. We reconstruct the spatiotemporal emission distribution from multi-detector Prompt Gamma Timing (PGT) data. Solving this inverse problem relies on an accurate  model of the prompt-gamma transport and detection including explicitly the dependencies on the time of emission and detection.
Our previous work relied on Monte-Carlo (MC) based system models. The tradeoff between computational resources and statistical noise in the system model prohibits studies of new detector arrangements and beam scanning scenarios. Therefore, we propose here an analytical system model to speed up recalculations for new beam positions and to avoid statistical noise in the model.
We evaluated the model for the MERLINO multi-detector-PGT prototype. Comparisons between the analytical model and a MC-based reference showed excellent agreement for single-detector setups. When several detectors were placed close together and partially obstructed each other, intercrystal scatter led to differences of up to \SI[detect-all=true]{10}{\percent} between the analytical and MC-based model. Nevertheless, when evaluating the performance in reconstructing the spatiotemporal distribution and estimating the stopping power, no significant difference between the models was observed. Hence, the procedure proved robust against the small inaccuracies of the model for the tested scenarios. The model calculation time was reduced by 1500 times, now enabling many new studies for PGT-based systems.
\end{abstract}


\newacro{cpt}[CPT]{Charged Particle Therapy}
\newacro{esp}[eSP]{Electronic Stopping Power}
\newacro{cog}[COG]{Center of Gravity}
\newacroplural{cog}[COGs]{Centers of Gravity}
\newacro{pg}[PG]{Prompt Gamma}
\newacro{pgt}[PGT]{Prompt Gamma Timing}
\newacro{pgi}[PGI]{Prompt Gamma Imaging}
\newacro{fluka}[FLUKA]{FLUktuierende KAskade}
\newacro{fom}[FoM]{Figure of Merit}
\newacroplural{fom}[FoMs]{Figures of Merit}
\newacro{mlem}[ML-EM]{Maximum-Likelihood Expectation-Maximization}
\newacro{ml}[ML]{Maximum Likelihood}
\newacro{serpgt}[SER-PGT]{Spatiotemporal Emission Reconstruction from \acl{pgt}}
\newacro{pgti}[PGTI]{Prompt Gamma Time Imaging}
\newacro{fov}[FOV]{Field of View}
\newacro{spr}[SPR]{Stopping Power Ratio}
\newacro{mre}[MRE]{Mean Relative Error}
\newacro{pgtspe}[PGT-SPE]{Prompt-Gamma-Timing-based Stopping Power Estimation}
\newacro{tof}[TOF]{Time of Flight}
\newacro{drf}[DRF]{Detector Response Function}
\newacro{mc}[MC]{Monte-Carlo}
\newacro{asm}[aSM]{Analytical System Model}
\newacro{mcsm}[MCSM]{Monte-Carlo-based System Model}
\newacro{pdf}[PDF]{Probability Density Function}
\newacro{pet}[PET]{Positron Emission Tomography}

\section{Introduction}
\label{sec:introduction}
\IEEEPARstart{N}{ovel} imaging concepts based on inverse problem solving require the development of appropriate models for the corresponding direct (or forward) problem. In this work, we present the first analytical model of a recently introduced approach for in-vivo verification of particle therapy, namely \ac{serpgt}. \Ac{serpgt} aims to retrieve the spatiotemporal distribution of the secondary gamma photons emitted from the patient during the treatment \cite{Pennazio_2022} as a first step towards inferring the stopping power of the tissues traversed by the therapeutic particle beams \cite{ferrero2022frontiers}.

In contrast to the widely used gamma-based radiotherapy, where clinical patient-specific verification methods are well established, \ac{cpt} still lacks a clinically implemented solution. Several approaches are currently under investigation, yet none has so far reached routine clinical practice \cite{parodi2019latestinvivo,gianoli2024verification,krann2024technologiecaldev}. The lack of in-vivo verification methods in \ac{cpt} is of particular concern, since any mismatch between the dose treatment plan and the actual dose distribution can lead to drastic consequences for both tumor control and healthy tissue sparing. This is due to the nature of the dose delivered by charged particles like protons or heavier ions in matter, which is characterized by a highly localized energy deposition followed by a steep spatial dose gradient.

Many of the approaches proposed for in-vivo \ac{cpt} verification rely on the detection of the so-called \ac{pg} radiation and its correlation with the dose distribution and the particle range; \ac{pg} photons originate directly from nuclear reactions that occur almost instantaneously when charged particles interact with the atomic nuclei in the tissue, and thus carry information linking their emission position and time with the motion of the beam particles in the patient.

\Ac{serpgt} expands the concept of \ac{pgt}. \Ac{pgt} measures the time difference $t_D$ between incoming beam particles crossing a reference plane and the detection of prompt-gamma photons; the histogrammed distribution of detection times is called the \ac{pgt} spectrum \cite{Golnik_2014,hueso2015first}. Changes in the tissues traversed by the beam particles translate into changes in the \ac{pgt} spectrum, allowing mismatches between the expected and the actual particle range to be inferred through the analysis of the \ac{pgt} spectra \cite{Werner_2019,schellhammer2022multivariate,KIESLICH2025machinelearning_temporal_spectral}. Based on \ac{pgt}, the more recent \ac{pgti} \cite{Jacquet_2021} aims to reconstruct the particle range using a geometrical model of the photon transport assuming point-like detectors and a precalculated model of the beam particle motion. Our approach \ac{serpgt} goes one step beyond, as it also incorporates the temporal dimension of the prompt gamma emission in addition to the spatial one thanks to the strategic placement of multiple \ac{pgt} detectors \cite{Pennazio_2022}. The resulting spatiotemporal distribution allows the study of the motion of the charged particles, revealing the stopping power of the traversed materials \cite{ferrero2022frontiers}. 
The multiple \ac{pgt} spectra are equivalent to the different views provided by tomographic systems. As in emission tomography, the measurement process can be described by a system of linear equations. The link between the expected measured data and the unknown spatiotemporal distribution of the \acp{pg} is given by the system response model. An accurate characterization of this model is therefore essential for reliable reconstruction.

For our first studies, we used Monte Carlo simulations to calculate the system model. This approach can provide accurate models, but at the cost of long computing times, introduction of statistical noise and limited flexibility \cite{IRIARTE201630,Qi_2005,rafecas2004noiseinSM}. To overcome these limitations and enable further extensions of \ac{serpgt}, we have developed, for the first time, a probabilistic analytical model describing photon transport and detection in \ac{pgt}. The new model has been exhaustively compared with the corresponding Monte-Carlo based approach at various levels, including the effects on the estimation of the beam range and the stopping power.

\begin{figure}
    \centering
    \begin{tikzpicture}[scale=1]
 
    \filldraw[fill=orange, fill opacity=0.2] (-1.5,0.5) -- ++(3,0) -- ++ (0,-1) -- ++(-3,0) -- cycle;
    \shade[right color=teal!30,left color=teal, opacity = .75, xshift=-1cm] (2,-.15) to[out = 0, in = 0] ++(0,.3) -- (-2,0.1) node[above right, xshift=-.4cm] (proton) 
    {beam}
    -- ++(0,-0.1) -- cycle;

    \draw[ultra thick] (-2.3,-0.2) node[below]{$t_0$} -- ++(0,.5);

    \filldraw[fill=green, fill opacity=0.2] (-0.1905,2) -- ++(0.381,0) -- ++(0,0.381) -- ++(-0.381,0) --cycle;
    \filldraw[fill=green, fill opacity=0.2, rotate=12.86] (-0.1905,2) -- ++(0.381,0) -- ++(0,0.381) -- ++(-0.381,0) --cycle;
    \filldraw[fill=green, fill opacity=0.2, rotate=25.71] (-0.1905,2) node[above, opacity=1] {\hspace{.5em}$t_D$} -- ++(0.381,0) -- ++(0,0.381) -- ++(-0.381,0) --cycle;
    \filldraw[fill=green, fill opacity=0.2, rotate=38.57] (-0.1905,2) -- ++(0.381,0) -- ++(0,0.381) -- ++(-0.381,0) --cycle;
    \filldraw[fill=green, fill opacity=0.2, rotate=51.43] (-0.1905,2) -- ++(0.381,0) -- ++(0,0.381) -- ++(-0.381,0) --cycle;
    \filldraw[fill=green, fill opacity=0.2, rotate=64.29] (-0.1905,2) -- ++(0.381,0) -- ++(0,0.381) -- ++(-0.381,0) --cycle;
    
    \filldraw[fill=green, fill opacity=0.2, rotate=-12.86] (-0.1905,2) -- ++(0.381,0) -- ++(0,0.381) -- ++(-0.381,0) --cycle;		
    \filldraw[fill=green, fill opacity=0.2, rotate=-25.71] (-0.1905,2) -- ++(0.381,0) -- ++(0,0.381) -- ++(-0.381,0) --cycle;
    \filldraw[fill=green, fill opacity=0.2, rotate=-38.57] (-0.1905,2) -- ++(0.381,0) node[above, opacity=1] {$t_D$} -- ++(0,0.381) -- ++(-0.381,0) --cycle;
    \filldraw[fill=green, fill opacity=0.2, rotate=-51.43] (-0.1905,2) -- ++(0.381,0) -- ++(0,0.381) -- ++(-0.381,0) --cycle;
    \filldraw[fill=green, fill opacity=0.2, rotate=-64.29] (-0.1905,2) -- ++(0.381,0) -- ++(0,0.381) -- ++(-0.381,0) --cycle;
    	
	\draw[decorate, decoration={snake}, draw=red, thick, ->] (0.9,0) -- (0.2,-.6);
	\draw[decorate, decoration={snake}, draw=red, thick, ->] (.1,0.02) node[yshift=-.2cm] {$(\point{P},t_\point{P})$}  -- (1.2, 1.6);
	\draw[decorate, decoration={snake}, draw=red, thick, ->] (-.3,-0.01) -- node[right] {\textcolor{red}{$t_\gamma$}} (-1, 1.75);
	\node[below left, inner sep=0, text opacity=1] at(1.5,.5) {target};

\end{tikzpicture}
    \caption{The multi-detector \acs*{pgt} measurement principle: a beam particle passes a reference plane (black line) at time \num{0}. After time $t_\point{P}$ the particle causes a \ac{pg} emission at position $\point{P}$. The photon takes time $t_\gamma$ to travel from $\point{P}$ to a detector (green) outside the irradiated target (orange), where it is detected at time ${t_D=t_\point{P}+t_\gamma}$.}
    \label{fig:pgt_principle}
\end{figure}
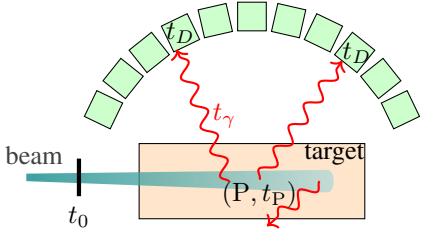

\section{Prompt gamma timing and spatiotemporal emission reconstruction}
The principle of \ac{pgt} with multiple detectors is visualized in \cref{fig:pgt_principle}. The basis are coincidence time measurements between a beam particle detector and secondary photon detectors surrounding the patient. Histograms of the time differences between beam and secondary particle detection, $t_D$, are called \ac{pgt} spectra.
$t_D$ includes the time the particle travels until it undergoes a nuclear reaction leading to a \ac{pg} emission, $t_\point{P}$, and the time it takes the photon to travel to the detector, $t_\gamma$. 
The latter ($t_\gamma$) depends on the emission position $\point{P}$ relative to the detector. 
Therefore, each \ac{pgt} spectrum contains complementary information about the \ac{pg} emission distribution in time and space. Our method \ac{serpgt}  exploits these facts to extract the spatiotemporal emission distribution of \ac{pg} photons from the \ac{pgt} spectra from multiple detectors \cite{Pennazio_2022,Ferrero2023IEEEexpReco}.

Central to \ac{serpgt} is the modeling of the forward problem, i.e., predicting the expected value $\mathbf{\bar{Y}}$ of the measurement for a given emission distribution $\mathbf{X}$. The measurement vector, $\mathbf{Y}=\{y_{d,n}\}$, consists of the \ac{pgt} spectra from  all detectors.
In \ac{serpgt} the discretized emission distribution, $\mathbf{X}=\{x_{j,p}\}$, can have up to three spatial dimensions in addition to the temporal dimension. $\mathbf{H}$ is the so-called system matrix, whose elements ${h_{d,n,j,p}=\prob(d,n|j,p)}$  express the conditional probability of a detection in detector $d$ within time interval $n$, given an emission within voxel $j$ during time interval $p$.
In the absence of background, the calculation of the expected measurement, $\mathbf{\bar{Y}}$, using $\mathbf{X}$ and $\mathbf{H}$ is expressed as a system of linear equations:
\begin{equation}
    \mathbf{\bar{Y}}=\mathbf{H}\mathbf{X}, \text{ or equivalently } \bar{y}_{d,n}=\sum_j \sum_p h_{d,n,j,p}x_{j,p}.
    \label{eq:direct_problem}
\end{equation}
The inverse problem consists of estimating the emission distribution, $\mathbf{X}$, from a given measurement, $\mathbf{Y}$. 
The accuracy of the reconstructed distribution depends, among other things, on how accurately  $\mathbf{H}$ describes the emission and detection processes involved in \ac{pgt}.

Assuming that the data are Poisson distributed, we can construct the corresponding Poisson likelihood function and use the \ac{ml} criterion to find the sought distribution.  In this work, \ac{mlem} is employed:
\begin{equation}
    \mathbf{\hat{X}}^{(k+1)}=\frac{\mathbf{\hat{X}}^{(k)}}{\mathbf{S}}\mathbf{H}^{\operatorname{T}}\frac{\mathbf{Y}}{\mathbf{H}\mathbf{\hat{X}}^{(k)}}
\end{equation}
with $\mathbf{\hat{X}}^{(k)}$ the estimated emission distribution after $k$ iterations and sensitivity $\mathbf{S}$ \cite{Pennazio_2022}.

The relationship between $\point{P}$ and $t_\point{P}$ encoded in $\mathbf{X}$ indirectly describes the beam particle motion, which is closely linked to the stopping power of the traversed tissue. \Ac{serpgt} does not require any assumptions about the dependencies between $\point{P}$ and $t_\point{P}$, but reveals them through the reconstructed distribution. The stopping power is a crucial parameter in treatment planning; our method thus strives to confirm its values during each treatment fraction. To this aim, we have proposed to combine \ac{serpgt} with motion models and non-linear regression to obtain the stopping power estimates \cite{ferrero2022frontiers,Werner_2024}. Consequently, the accuracy of the \ac{serpgt} output, $\mathbf{\hat{X}}$, has a direct impact on the accuracy of the stopping power estimates.

\section{System models}
Here we describe the two methods to calculate the system matrix elements.
Effects like the finite time resolution of the detectors are added in postprocessing to the models as a convolution along the detection time dimension with the normalized distribution of possible deviations in the time measurement. In the following sections, \textit{detector} refers to secondary particle detectors and not the primary particle detector. Details of the detector system used for comparing the two models are described in \cref{sec:evaluation}.

\subsection{\acf*{mcsm}}
We simulated a known number of photons being emitted along the spatial dimension of the \ac{fov}.
The origin of the photons were randomly selected following a uniform distribution.
The emission directions were randomly sampled and isotropically distributed. All photons were emitted at time interval center ${t_p}$, i.e., the width of the emission time bin $\Delta t_\point{P}$ was not considered. In the simulated volume, everything but the scintillation crystals of the photon detectors was vacuum. 

For each energy deposition within a detector, we recorded the detector index $d$ along with the photon emission position, $\point{P}$, detection time, $t_D$, and deposited energy, $E_D$. Next, we counted the number of detections separately for each combination of detector $d$, emission position interval $j$, emission time interval $p$, and detection time interval $n$. The elements of the system matrix, $\mathbf{H}$, were estimated as
\begin{align}
    h_{d,n,j,p}^\text{\acs*{mcsm}}=\frac{N_{d,n,j,p}}{N_{j,p}},
\end{align}
i.e., the ratio of the number of events in detector $d$ within detection time interval $n$ after emission in emission position voxel $j$ and emission time interval $p$, ($N_{d,n,j,p}$), and the number of photons emitted within intervals $j$ and $p$, ($N_{j,p}$).

\subsection{\Acf*{asm}}
\begin{table}[!t]
    \centering
    \caption{Mathematical symbols and short descriptions.}
    \label{tab:symbol_descriptions}
    \begin{tabularx}{\columnwidth}{lX}
         \toprule
         Symbol& Description \\
         \midrule
         $h_{d,n,j,p}$ & system matrix element \\
         $d$ & detector index \\
         $n$ & index of detection time interval \\
         $j$ & index of emission position interval \\
         $p$ & index of emission time interval \\
         $i$ & index of traversed materials \\
         $t_n$, $\point{P}_j$, $t_p$ & centers of intervals $n$, $j$, and $p$\\[.5em]
         $\Delta x$, $\Delta y$, $\Delta z$ & spatial sizes of PG emission bins \\
         $\Delta t_\point{P}$ & temporal size of PG emission bins \\      
         $\Omega$ & emission direction as an infinitesimal solid angle \\
         $\Delta t_D$ & temporal size of detection time interval \\
         
         $z$ & position along beam axis relative to isocenter \\
         $\point{P}$, $t_\point{P}$ & point and time of PG emission \\
         $\dd V_\point{P}$ & volume integrand related to $\point{P}$ \\
         $\theta$ & polar angle of emission direction\\
         $\varphi$ & azimuthal angle of emission direction\\
         $\Delta \theta, \Delta \varphi$ & angle discretization step sizes \\[.5em]
         
         $t_D$ & timestamp of detection event \\
         $\point{Q}$ & point of entry into detector for given ray \\
         $l$ & distance from detector entry $\point{Q}$ \\
         $l_1$, $l_2$ & depths within the crystal at the start and at the end of the detection time interval, respectively\\
         $L_1$ & path length in detector before the time interval \\
         $L_2$ & path length in detector during time interval \\
         $L_d$ &  intersection length between ray and detector $d$\\[.5em]
         
         $c$ & speed of light\\
         $\mu$ & linear attenuation coefficient \\
         $\mu_D$ & probability of detection along path element $\dd l$ \\
         $E_D$ & measured energy deposition \\
         $a_{d,z,\theta,\varphi}$ & attenuation between $\point{P}$ and $\point{Q}$ \\
         \bottomrule
    \end{tabularx}
\end{table}

All mathematical symbols of this section are listed in \cref{tab:symbol_descriptions}. 
Throughout this section the shorthand 
\begin{equation}
    [a\pm\Delta a]:=[a-\Delta a,a+\Delta a]
    \label{eq:def_shorthand}
\end{equation}
will be used. If $a$ is a  three-dimensional position, \cref{eq:def_shorthand} shall apply to each dimension separately, i.e., in that case ${[a\pm\Delta a]}$ describes a cuboid centered in $a$. $\diff{\point{AB}}=\diff{\point{BA}}$ is the distance between two points $\point{A}$ and $\point{B}$.
A ray is defined as the line defined by a three-dimensional start position and direction.

First, we inspect the conditions necessary for a valid \ac{pgt} detection event given a specified emission.
Interactions leading to a detected event can only take place within the volume of the detector $d$.
Consequently, the depth of the interaction within the detector, $l$, must be positive and not exceed the overall intersection length $L_d(\point{P},\Omega)$ of the given ray with detector $d$. If the ray does not cross the detector volume, the length of the interval and, thereby, the associated probability of detection is zero.
The detection time, $t_D$, must lie in the time interval $n$, i.e., $t_D\in[t_n\pm\Delta t_D/2]$. Since the photon is moving at a finite speed, we must consider which combinations of detection position and time are possible. Let $\point{Q}(\point{P},\Omega,d)$ be the point at which the ray first enters the detector. The condition 
\begin{equation}
c(t_D-t_\point{P})= l + \diff{\point{Q}\point{P}} \label{eq:t_d_to_l}
\end{equation}
derives from $t_D=t_\point{P}+t_\gamma$, with 
$t_\gamma=(l + \diff{\point{Q}\point{P}})/c$. This ensures that the distance from emission to the point of detection, ${l + \diff{\point{Q}\point{P}}}$, is consistent with the photon traveling at the speed of light $c$ in a straight line for time $t_D-t_\point{P}$. The energy of \acp{pg} is hundreds of \si{\kilo\electronvolt} and above. Therefore, we assume the speed of light in vacuum is a reasonable estimate for the photons speed in the irradiated target and detector material. Lastly, consider that we are searching for the detection probability preconditioned on an emission in a specified voxel $j$ and time interval $p$. Therefore, an emission inside these intervals must be assumed for the following calculations. All these conditions can be summarized as
\begin{align}
    \prob(d,n|j,p) = \prob(&l(\point{P},\Omega,d)\in[0,L_d(\point{P},\Omega)] \nonumber\\
    &\quad\text{and } t_D\in[t_n\pm\Delta t_D/2] \nonumber\\
    &\quad\text{and } c(t_D-t_\point{P})= l + \diff{\point{Q}\point{P}} \nonumber\\
    &\text{given }\point{P}\in[\point{P}_j\pm\Delta\point{P}/2] \nonumber\\
    &\quad\text{and }t_\point{P}\in[t_p\pm\Delta t_\point{P}/2]).
    \label{eq:P_as_text}
\end{align}

\begin{figure}[!t]
    \centering
    \begin{subfigure}{.49\columnwidth}
        \centering
        \begin{tikzpicture}[scale=0.7]

    \coordinate (emission) at (0,0);

    \fill[opacity=0.25, blue]{} (emission) ++ (-12:3.5) arc(-12:-65:3.5) -- (-48:4.25) arc(-48:-12:4.25) -- cycle;
    \fill[opacity=0.25, red]{} (emission) -- ++ (-12:3.5) arc(-12:-65:3.5) -- cycle;

    \fill[fill=green, fill opacity=0.2] (2,-1) -- ++(0,-2) -- ++(1.5, 0) -- ++(0,2) -- cycle;

    \draw[gray, dashed] (emission) -- ++ (-25:4.75);
    \draw[red] (emission) +(-25:2.38) -- node[midway, below, xshift=0.1cm, yshift=0.05cm]{$L_1$} ++(-25:3.5);
    \draw[blue] (emission) ++ (-25:3.5) -- ++(-25:.37);

    \draw[gray, dashed] (emission) -- ++ (-45:4.75);
    \draw[red] (emission) +(-45:2.825) -- ++(-45:3.5);
    \draw[blue] (emission) ++ (-45:3.5) -- node[midway, above, xshift=0.2cm]{$L_2$}++(-45:.75);

    \draw[gray, dashed] (emission) -- ++ (-35:4.75);
    \draw[red] (emission) +(-35:2.45) -- ++(-35:3.5);
    \draw[blue] (emission) ++ (-35:3.5) -- ++(-35:.75);
    
    \draw[] (emission) ++ (-10:3.5) node[above] {$l_{1}$} arc(-10:-65:3.5);
    \draw[] (emission) ++ (-10:4.25) node[above] {$l_{2}$} arc(-10:-48:4.25);

    \draw (2,-1) -- ++(0,-2) -- ++(1.5, 0) -- ++(0,2) -- cycle;

    \draw[-latex] (-1,0) -- (4.8,0) node[right] {$z$}; 
    \filldraw[black] (emission) circle (2pt) node[above] {$(\point{P},t_P)$};
    \filldraw[black] (emission) +(-25:2.38) circle (2pt) node[above] {$\point{Q}_\text{i}$};
    \filldraw[black] (emission) +(-45:2.825) circle (2pt) node[left] {$\point{Q}_\text{iii}$};
    \filldraw[black] (emission) +(-35:2.45) circle (2pt) node[left] {$\point{Q}_\text{ii}$};
\end{tikzpicture}
        \caption{}
        \label{sfig:model_principle}
    \end{subfigure}
    \begin{subfigure}{.49\columnwidth}
        \centering 
        \begin{tikzpicture}[scale=0.9]
    \filldraw[fill=green, fill opacity=0.2] (0,3.2) -- (0,.3)  -- (2,.3) -- (2,3.2) --cycle;

    \draw[>-*, dotted] (-.5,3) node[left] {1)\hspace{0.1cm}\ } -- (0,3) edge[blue, -, solid, thick] (2,3) -- (2.5,3);
    \draw[>-*, dotted] (-.5,2.5) node[left] {2)\hspace{0.1cm}\ } -- (0,2.5) edge[blue, -, solid, thick] (1.5,2.5) -- (1.665,2.5);
    \draw[>-*, dotted] (-.5,2) node[left] {3)\hspace{0.1cm}\ } -- (-0.05,2);

    \draw[red, thick] (0,1.5) -- (.5,1.5);
    \draw[blue, -, solid, thick] (0.5,1.5) -- (2,1.5);
    \draw[dotted, >-*] (0.5,1.5) -- (2.5,1.5);
    \node[left, black] at(-.5,1.5) {4)\hspace{0.1cm}\ };
    
    \draw[red, thick] (0,1) -- (.5,1);
    \draw[blue, -, solid, thick] (.5,1) -- (1.5,1);
    \draw[dotted, >-*] (0.5,1) -- (1.65,1);
    \node[left, black] at(-.5,1) {5)\hspace{0.1cm}\ };

    \draw[>-*, dotted] (2.1,.5) -- (2.5,.5);
    \node[left, black] at(-.5,.5) {6)\hspace{0.1cm}\ };

\end{tikzpicture}
        \caption{}
        \label{sfig:startendcombinations}
    \end{subfigure}
    \caption{(a) 2D representation of a detector (box) and three emission directions (dashed lines) with detector entry points $\point{Q}_\text{i}$, $\point{Q}_\text{ii}$, and $\point{Q}_\text{iii}$ and path lengths in the detector before ($L_1$, red) and 
    within detection interval $[l_1, l_2]$
    ($L_2$, blue).
    (b) Sections of attenuation within the detector before detection time interval $n$ ($L_1$, red) and detection ($L_2$, blue) for all relevant combinations of the start ($>$) and end (\textbullet) of the detection time interval relative to the detector volume (green).
    }
\end{figure}
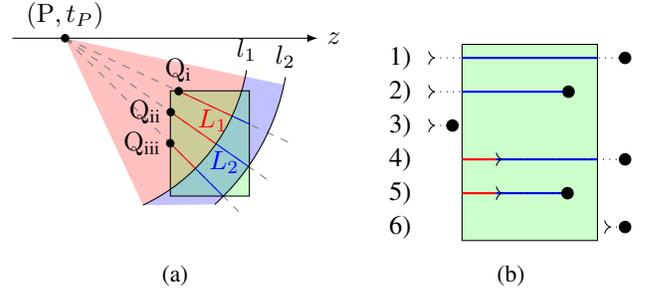

To calculate $\prob(d,n|j,p)$, we derive $\dd \prob_{\point{P},t_\point{P},t_D,l,\Omega}(d,n|j,p)$, which is the differential probability of detection at time $t_D$ at depth $l$ in detector $d$ given an emission at time $t_\point{P}$ and position $\point{P}$ in direction $\Omega$. The emission position and direction defines the assumed photon path after emission (ray) as sketched in \cref{sfig:model_principle}. Once fully described, we integrate the differential probability $\dd \prob_{\point{P},t_\point{P},t_D,l,\Omega}(d,n|j,p)$ over each of the free variables $\point{P}$, $t_\point{P}$, $t_D$, $l$, and $\Omega$, one at a time. This yields the total probability $\prob(d,n|j,p)$ for the given combination of emission and detection intervals.

The differential probability is decomposed into factors, each representing conditions outlined in \cref{eq:P_as_text}.
To describe the voxel, time interval and detector boundaries, the indicator function
\begin{equation}
    \mathds{1}_{[a,b]}(\tau)=\begin{cases}
        1,&\text{if}~ \tau \in [a,b];\\
        0,&\text{otherwise}.
    \end{cases}
\end{equation}
is used. The function 
\begin{equation}
    \mathds{1}_j(\point{P})=\mathds{1}_{[x_j\pm\Delta x/2]}(x_P)\,\mathds{1}_{[y_j\pm\Delta y/2]}(y_P)\,\mathds{1}_{[z_j\pm\Delta z/2]}(z_P)
    \label{eq:pdf_voxel}
\end{equation}
indicates whether the emission position, $\point{P}$, lies within voxel $j$ with voxel center $\point{P}_j$. Similarly, the intervals for the emission and detection times, as well as the valid interaction depths within the detector are indicated as follows:
\begin{align}
    \mathds{1}_p(t_\point{P})&=\mathds{1}_{[t_p\pm\Delta t_\point{P} / 2]}(t_\point{P}),\\
    \mathds{1}_n(t_D)&=\mathds{1}_{[t_n\pm\Delta t_D / 2]}(t_D),\\
    \mathds{1}_{\point{P},\Omega,d}(l) &=\mathds{1}_{[0,L_d(\point{P},\Omega)]}(l),\label{eq:indicator_detector}
\end{align}
where $t_p$ and $t_n$ are the time interval centers, respectively, and $L_d$ is the maximum possible path length of a photon in detector $d$ based on the emission position and direction.
Additionally, all possible emission directions (characterized by the emission solid angle $\Omega$) need to be considered. Here we assume the \ac{pg} emission to be isotropic, altough the non-isotropy of \ac{pg} emission could be modeled as a weighting factor dependent on $\Omega$.
The indicator functions for $\point{P}$, $\Omega$ and $t_\point{P}$ are normalized with regard to their interval sizes $\Delta x$, $\Delta y$, $\Delta z$, $4\pi$ and $\Delta t_\point{P}$, respectively, 
since we calculate the detection probability conditioned on an emission in spatiotemporal voxel $(j,p)$ with arbitrary direction. 

The probability of a \ac{pg} to be detected at a certain depth $l$ requires that the photon first reaches that point (survival probability) after which it interacts over an infinitesimal path length $\dd l$. According to the Beer-Lambert law for a material with attenuation coefficient $\mu$, we describe the survival probability as
\begin{equation}
    f_d(l) = e^{-\mu l}.
    \label{eq:beer_lambert}
\end{equation}
\Cref{ssec:results_timestamp} shows that build-up effects can be ignored when translating detection times into interaction positions using \cref{eq:t_d_to_l}. 
Additionally, we consider $\mu_D$, which is the probability of detection per infinitesimal path element $\dd l$. 
It may contain any effect that is independent of $\point{P}$, $\Omega$, $t_\point{P}$, and $t_D$. One example is the probability of a given interaction to deposit enough energy into the detector to be counted as a measurement event. This may include a detector-dependent efficiency calibration; $\mu_D$ will be equal to or smaller than $\mu$, since some interactions are not counted in the \ac{pgt} spectra, based on event selection criteria.

Combining the conditions described in \crefrange{eq:pdf_voxel}{eq:beer_lambert} yields 
\begin{align}
    \dd \prob_{\point{P},t_\point{P},t_D,l,\Omega}&(d,n|j,p)
    \nonumber\\=&
    \dd V_\point{P} \frac{\operatorname{\mathds{1}}_j(\point{P})}{\Delta x \Delta y \Delta z}
    \frac{\dd\Omega}{4\pi}\,
    \dd t_\point{P} \frac{\operatorname{\mathds{1}}_{p}(t_\point{P})}{\Delta t_\point{P}}
    \nonumber\\&
    \dd t_D \operatorname{\mathds{1}}_{n}(t_D)\,
    \dd l\, \mu_D\, e^{-\mu l} \operatorname{\mathds{1}}_{\point{P},\Omega,d}(l)
    \nonumber\\&
    \delta\left(t_D-\left(l+\diff{\point{Q}\point{P}}\right)/c-t_\point{P}\right);
    \label{eq:diffProb}
\end{align}
the first line describes the probability of emission within the correct voxel and time interval in all possible directions. The differential volume of $\point{P}$ is $\dd V_\point{P}$.
The second line deals with detection probabilities, both in terms of detection time and locations within the detector. Lastly, we use ${\delta\left(t_D-\left(l+\diff{\point{Q}\point{P}}\right)/c-t_\point{P}\right)}$ to only include combinations that fulfill \cref{eq:t_d_to_l}.

First, we integrate over $t_D$, using the sifting property of the $\delta$ distribution:
\begin{align}
    \dd \prob&_{\point{P},t_\point{P},l,\Omega}(d,n|j,p) \nonumber\\
    =&\int_{t_D\in\mathbb{R}}\dd \prob_{\point{P},t_\point{P},t_D,l,\Omega}(d,n|j,p) \nonumber \\
    =&
    \dd V_\point{P} \frac{\operatorname{\mathds{1}}_j(\point{P})}{\Delta x \Delta y \Delta z}
    \frac{\dd\Omega}{4\pi}\,
    \dd t_\point{P} \frac{\operatorname{\mathds{1}}_{p}(t_\point{P})}{\Delta t_\point{P}}
    \operatorname{\mathds{1}}_{n}\left(\left(l+\diff{\point{Q}\point{P}}\right)/c+t_\point{P}\right)\, \nonumber\\&
    \dd l\, \mu_D\, e^{-\mu l} \operatorname{\mathds{1}}_{\point{P},\Omega,d}(l)
    \label{eq:dP_P_tp_l_omega}
\end{align}
Next, we integrate this result over  $t_\point{P}$ to obtain $\dd \prob_{\point{P},l,\Omega}(d,n,j,p)$. The \cref{sec:app_convolution} describes how this integral can be interpreted as a convolution of $\operatorname{\mathds{1}}_p$ and $\operatorname{\mathds{1}}_n$.
To simplify the calculation, we evaluate $\prob(d,n|j,p)$ only at the center of the emission time and position intervals, assuming that the change within the spatiotemporal voxel is negligible:
\begin{align}
\dd \prob&_{l,\Omega}(d,n|j,p) \approx  \lim_{\Delta x, \Delta y, \Delta z, \Delta t_\point{P} \to 0} \int_{\mathbb{R}^4} \dd \prob_{\point{P},t_\point{P},l,\Omega}(d,n|j,p).
\end{align}
Making use of
\begin{align}
    \lim_{\Delta x, \Delta y, \Delta z\to 0}\frac{\operatorname{\mathds{1}}_j(\point{P})}{\Delta x \Delta y \Delta z} &=\delta(x_P-x_j)\delta(y_P-y_j)\delta(z_P-z_j),
    \label{eq:lim_1_j}
\end{align}
and
\begin{align}
    \lim_{\Delta t_\point{P}\to 0}\frac{\operatorname{\mathds{1}}_p(t_\point{P})}{\Delta t_\point{P}}&=\delta(t_\point{P}-t_p)
    \label{eq:lim_1_p},
\end{align}
together with  the sifting property of  $\delta$, we obtain
\begin{align}
    \dd \prob&_{l,\Omega}(d,n|j,p) \nonumber\\
    &=\frac{\dd\Omega}{4\pi}\,\dd l\, \mu_D \, e^{-\mu l}
    \operatorname{\mathds{1}}_{\point{P}_j,\Omega,d}(l)\,
    \mathds{1}_n((l+\diff{\point{Q}\point{P}_j})/c+t_p).
\end{align}
By substituting the detection times in bin $n$, with distances within the detector via \cref{eq:t_d_to_l}, i.e., ${\left[l_1,l_2\right]=[c\left(t_n-t_p\pm\Delta t_D/2\right)-\diff{\point{Q}\point{P}_j}]}$, we rewrite  the last indicator function to have $l$ as an input
\begin{align}
    \mathds{1}_n((l+\diff{\point{Q}\point{P}_j})/c+t_\point{P}) = \mathds{1}_{[l_1,l_2]}(l).
\end{align}
Together with \cref{eq:indicator_detector} this results in
\begin{align}
    \dd \prob_{l,\Omega}(d,n|j,p) 
    =&\frac{\dd\Omega}{4\pi}\dd l\, \mu_D\, e^{-\mu l}\, \mathds{1}_{[0,L_d(\point{P}_j,\Omega)]}(l),
    \, \mathds{1}_{[l_1,l_2]}(l)
    \label{eq:dP_l_omega}
\end{align}
which contains two indicator functions with the same input variable.
Their multiplication yields another indicator function:
\begin{align}
    \mathds{1}_{[0,L_d(\point{P}_j,\Omega)]}(l) \, \mathds{1}_{[l_1,l_2]}(l) = \mathds{1}_{[L_1, L_1 + L_2]}(l).
    \label{eq:mult_indicators}
\end{align}
with
\begin{equation}
    L_1 = \max(0,l_1)
\end{equation}
and
\begin{equation}
    L_2 = \begin{cases}
        0,\hfill \text{if }l_1 \geq L_d\text{ or }l_2<0, \\
        \min(l_2,L_d) - L_1,\quad\text{otherwise}.
    \end{cases}
\end{equation}
To better understand this result, we examine the overlap of detector volume and detection time interval more closely.
\Cref{sfig:model_principle} depicts three possible photon trajectories emitted from a $\point{P}$ at time $t_p$ passing through the detector. Valid detection times can occur at distances between $l_1+\diff{\point{P}\point{Q}}$ and $l_2+\diff{\point{P}\point{Q}}$ from the photon origin $\point{P}$. This can be visualized as a hollow sphere centered on $\point{P}$, whose intersection with the detector volume contains all interaction positions, where a detection event in detector $d$ would be counted in time bin $n$. For a given ray, $L_1$ (red segment) represents the distance to be traveled through the detector before reaching the detection interval. The length of the intersection segment is $L_2$ (blue), which cannot be larger than the total intersection length between the detector volume and the ray, $L_d$.  
\Cref{sfig:startendcombinations} shows the six possible combinations of the intervals $[l_1,l_2]$ and $[0,L_d]$: For a given ray,
the detection time interval (blue region) either
\begin{enumerate}
    \item fully encloses the intersection of detector volume and ray,
    \item starts before  and ends within the detector volume,
    \item lies entirely between the point of emission and the detector,
    \item starts within the detector and ends behind the distal edge of the detector,
    \item starts and ends within the detector volume, or
    \item lies entirely behind the detector as viewed from the point of emission.
\end{enumerate}
For emission directions where detector $d$ is not crossed, ${L_d=0}$ implies ${L_2=0}$, i.e., an intersection length of $0$.
Using \cref{eq:mult_indicators}, the detection probability for a given emission direction reduces to
\begin{align}
    \dd \prob_{\Omega}(d,n|j,p)     
    &=\frac{\dd\Omega}{4\pi} \,  \mu_D\int_{L_1}^{L_1+L_2} \dd l\,  e^{-\mu l} \nonumber \\
    &= \frac{\dd\Omega} {4\pi} e^{-\mu L_1} (1 - e^{-\mu L_2}) \frac{\mu_D}{\mu},
\end{align}
i.e., four independent probabilities. $\frac{\dd\Omega} {4\pi}$ is the differential probability of emission in a given direction. This is followed by the probability of the photon reaching depth $L_1$ inside the detector $e^{-\mu L_1}$ and then the probability ${1 - e^{-\mu L_2}}$ of the photon depositing any energy in an appropriate part of the detector. Since some energy depositions may be too small or are rejected for other reasons, we finally have the probability $\frac{\mu_D}{\mu}$ of an energy deposition to be counted in the \ac{pgt} spectrum.

The final step is to integrate the differential probability over all emission directions to obtain the total probability. Using 
spherical coordinates,  the probability of detection within time bin $n$ in detector $d$ given an emission in $(j,p)$ is obtained through
\begin{equation}
    \prob(d,n|j,p) = \int_0^{2\pi}\dd\varphi\, \int_0^{\pi} \dd \theta \sin\theta \, \frac{e^{-\mu L_1}}{4\pi} \frac{\mu_D}{\mu}  (1 - e^{-\mu L_2}).
\end{equation}
For each direction, the entry point to the detector, $\point{Q}$, and the path length in the detector, $L_d$, must be calculated. Closed-form expressions for the two might not be found, depending on the detector geometry. Therefore, we approximate the probability by applying the rectangular rule and discretizing the solid angle with equally spaced increments of the azimuthal ($\varphi$) and polar ($\theta$) angles. The final form of a system matrix element is
\begin{equation}
    h_{d,n,j,p}^\text{\acs*{asm}}=\frac{\Delta\varphi\Delta\theta}{4\pi}\frac{\mu_D}{\mu}\sum_{m=0}^{\left\lfloor\frac{2\pi}{\Delta\varphi}\right\rfloor} \sum_{k=0}^{\left\lfloor{\frac{\pi}{\Delta \theta}}\right\rfloor} \sin\theta_k \, e^{-\mu L_1}  (1 - e^{-\mu L_2}).
\end{equation}

Attenuation between emission and detector entry can be modeled into $h_{d,n,j,p}^{\text{\acs*{asm}}}$ by adding the multiplicative term 
\begin{equation}
    a(d,\point{P},\Omega)=e^{-\sum_i \mu_{i}l(i,d,\point{P},\Omega)},\label{eq:inter_crystalattenuation}
\end{equation}
with $i$ being the index of traversed materials; $a$ depends on the emission position, $\point{P}$, and direction, $\Omega$, as well as detector index $d$. In this work, we have implemented the additional attenuation of other detectors. 
Attenuation in the phantoms or other surrounding materials was not considered. 

\section{Evaluation}\label{sec:evaluation}
All tests were executed with a spatiotemporal \ac{fov} extending from \SIrange{-20}{20}{\centi\meter} on the beam axis relative to the isocenter, and from \SIrange{0}{5}{\nano\second} with a discretization of ${\Delta z=\SI{0.25}{\centi\meter}}$, ${\Delta t_\point{P}=\SI{50}{\pico\second}}$, and ${\Delta t_D=\SI{50}{\pico\second}}$. 
For both models, only the system matrix elements for emission time bin ${p=0}$ needed to be calculated explicitly. 
Due to $t_D=t_\point{P}+t_\gamma$, a translation in emission time, $t_\point{P}$, shifts the detection time, $t_D$, by the same amount. Since $\Delta t_\point{P}$ and $\Delta t_D$ are equal, we obtained the system matrix values for $p\neq 0$ by shifting the system matrix elements obtained for $p=0$ in the detection-time dimension to match the desired emission time. 

Following our experimental setup \cite{Ferrero_2022JINST}, the detectors were modeled as LaBr$_3$ scintillation crystals with diameter and length of \SI{3.81}{\centi\meter} (unless stated otherwise) and density \SI{5.2}{\gram\per\centi\meter\cubed}. 
In this work, we calculated both models assuming mono-energetic \SI{4.4}{\mega\electronvolt} photons. For the \ac{asm}, the total attenuation coefficient for  LaBr$_3$,  ${\mu=\SI{17.71}{\per\meter}}$,  was taken from XCOM NIST\footnote{https://physics.nist.gov/PhysRefData/Star/Text/PSTAR.html}.

For the \ac{mcsm}, we simulated a total of \num{6E10} photons in FLUKA 2024.1.3 \cite{Ballarini2024FLUKA}, which took approximately 160 core-hours using an AMD\copyright\ Ryzen 7 pro 5750g. The number of primary photons was intentionally set very high to reduce the impact of statistical errors. 
The calculation of the \ac{asm} with $\Delta\theta=\Delta\varphi=\SI{1}{\degree}$ for 110 detectors took 6 core-minutes; a speedup of approximately 1500 times compared to the \ac{mcsm} implementation. 
In the following sections the \ac{mcsm} and \ac{asm} implementations are compared.

\begin{figure}
    \centering
    \begin{subfigure}[T]{.39\columnwidth}
        \centering
        \includegraphics[width=\columnwidth]{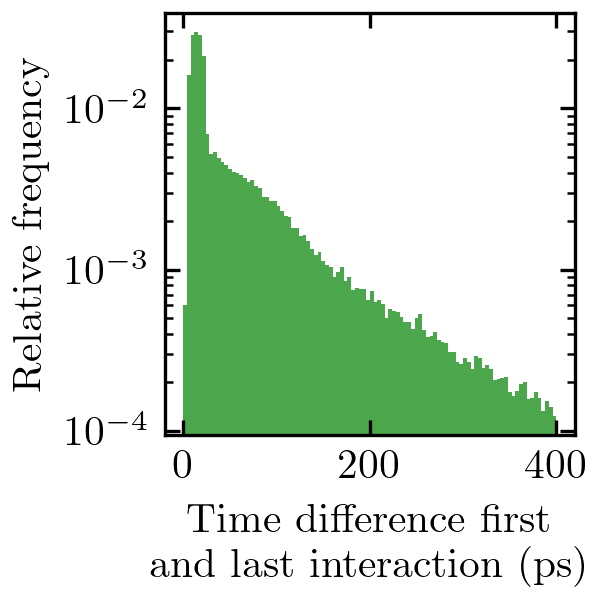}
        \caption{}
        \label{sfig:timediff_MC}
    \end{subfigure}
    \begin{subfigure}[T]{.59\columnwidth}
        \centering 
        \includegraphics[width=\columnwidth]{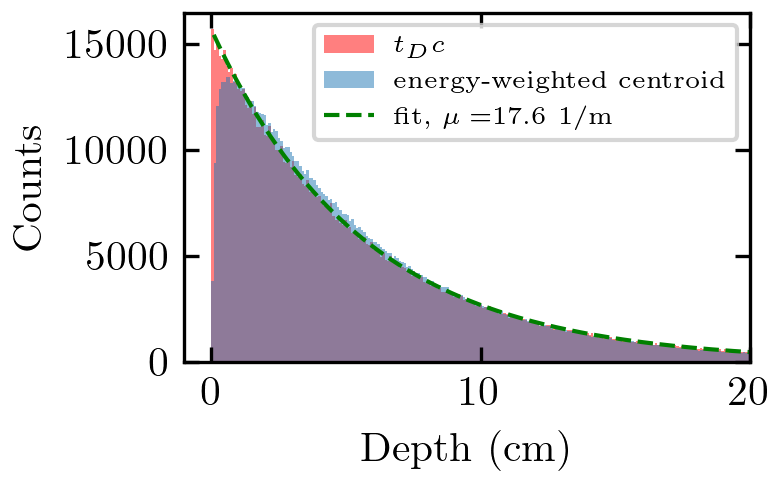}
        \caption{}
        \label{sfig:time_as_depth}
    \end{subfigure}
    \caption{(a) Time-difference spectrum between last and first energy deposition of a \SI{4.4}{\mega\electronvolt}-photon in the \acs*{mc} simulation. (b) Number of energy depositions over path length in LaBr$_3$ in a \acs*{mc}-simulation. The path length is calculated either using an energy-weighted centroid (blue) or the time after entry and speed of light (red). The fit of the time-based calculation to an exponential function is shown in green. }
    \label{fig:time_as_depth_and_DRF}
\end{figure}
\subsection{Timestamp calculation}\label{ssec:results_timestamp}
In the \ac{mcsm} model, all interactions of secondary particles from the same photon within a detector are aggregated to form a net detection event, i.e., their energy is summed, and only one timestamp is assigned. 
However, as illustrated in \Cref{sfig:timediff_MC}, the time difference between the first and last energy depositions related to one net event can be on the scale of hundreds of picoseconds. The median time difference was \SI{25}{\pico\second}. This distribution is hereafter termed \ac{drf}, as it also serves as a stand-in for any effect changing the relationship of the time of photon-detector interaction and the measured timestamp. Depending on the timing resolution, this effect needs to be taken into account when comparing the system models. For all following results, we chose the timestamp of the very first energy deposition related to the given primary photon as the overall detection time $t_D$.

We have also tested whether the translation of the traveled distance from the point of emission, $\point{P}$, into the photon time of flight via \cref{eq:t_d_to_l} is a reasonable approximation. For this purpose we simulated a \SI{100}{\centi\meter} long cylindrical crystal of LaBr$_3$ with diameter \SI{3.81}{\centi\meter}. \Cref{sfig:time_as_depth} shows a buildup effect in the first centimeters of the crystal, when calculating the interaction depth, $l$, as the energy-weighted centroid of energy depositions for the simulated data. The buildup effect is the initial rise of secondary particle fluence (and thereby counts), as the positions close to the borders of the crystal have less material surrounding them producing secondaries, which then travel some distance within the crystal. In contrast, when the timestamps from the simulation are translated into a distance, the results closely follow the Lambert-Beer Law: The fit to \cref{eq:beer_lambert} shows an attenuation coefficient of \SI{17.6}{\per\meter}, close to the value from XCOM NIST of \SI{17.71}{\per\meter}.

\begin{figure}
    \centering
    \begin{subfigure}{\columnwidth}
        \includegraphics[width=.8\linewidth]{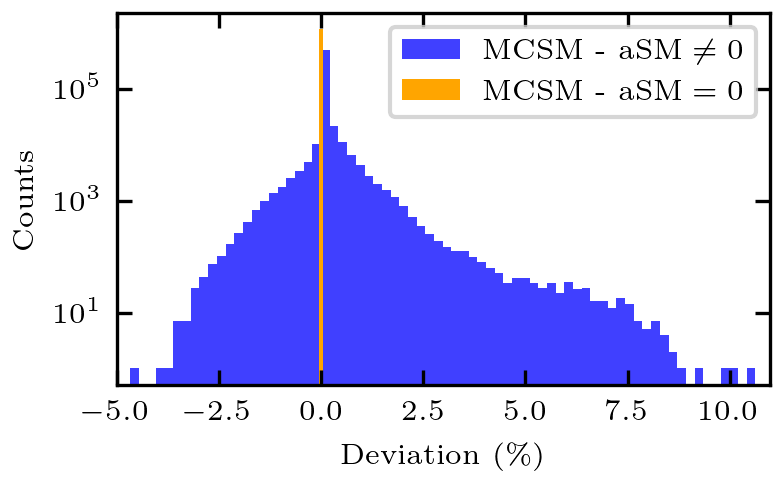}
        \caption{}
        \label{sfig:error_distribution}
    \end{subfigure}
    \begin{subfigure}{\columnwidth}
        \includegraphics[width=\linewidth]{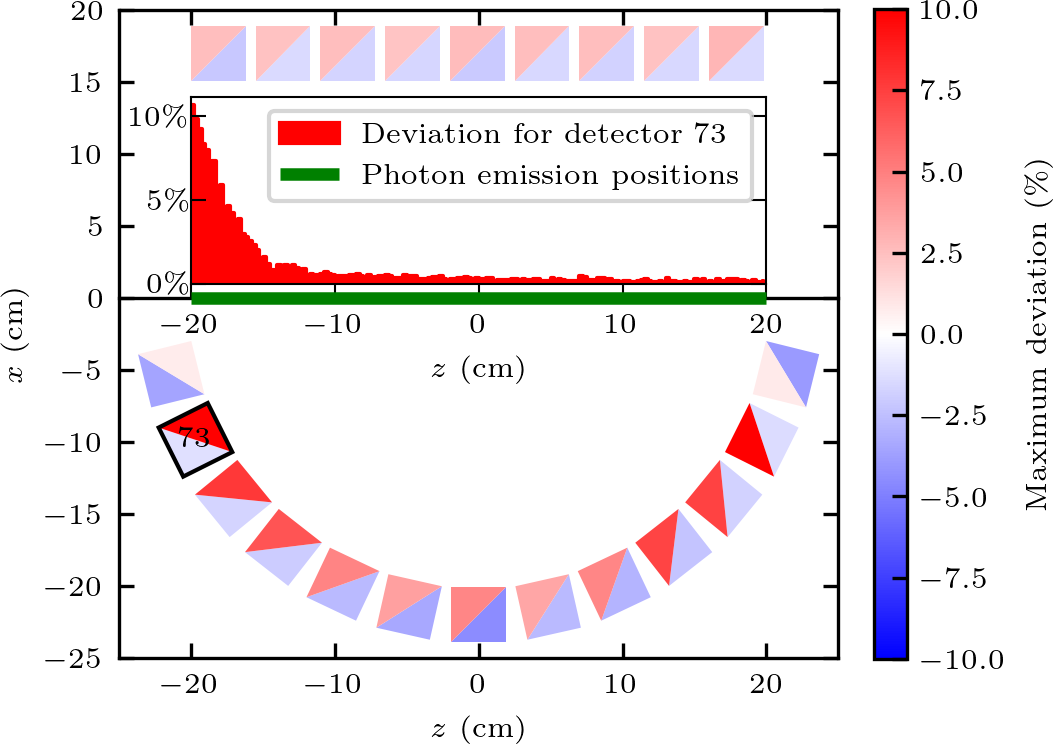}
        \caption{}
        \label{sfig:maxerror_slice}
    \end{subfigure}
    \caption{(a) Element-wise histogram of ${h_{d,n,j,p}^\text{\acs*{mcsm}}-h_{d,n,j,p}^\text{\acs*{asm}}}$ relative to $\max_{n,j,p}h_{d,n,j,p}^\text{\acs*{mcsm}}$ in detector $d$ with energy threshold \SI{1}{\mega\electronvolt}. (b) Maximum positive (red) and negative (blue) elementwise differences of ${h_{d,n,j,p}^\text{\acs*{mcsm}}-h_{d,n,j,p}^\text{\acs*{asm}}}$ for each detector in the xz-plane. Inset: Maximum difference depending on the emission positions for detector position 73.}
    \label{fig:max_deviations_asm_vs_mcsm}
\end{figure}
\begin{figure}
    \centering
    \begin{subfigure}{.8\columnwidth}
        \centering
        \includegraphics[width=\columnwidth]{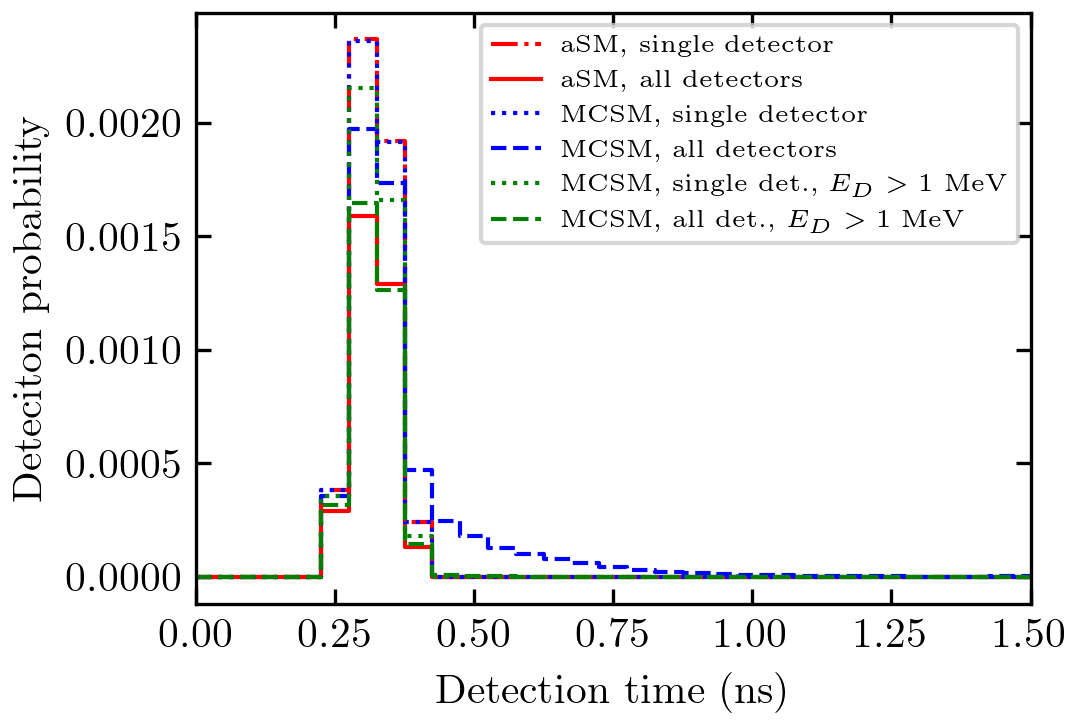}
        \caption{}
        \label{sfig:asm_vs_mc_-20cm}
    \end{subfigure}
    \begin{subfigure}{.8\columnwidth}
        \centering
        \includegraphics[width=\columnwidth]{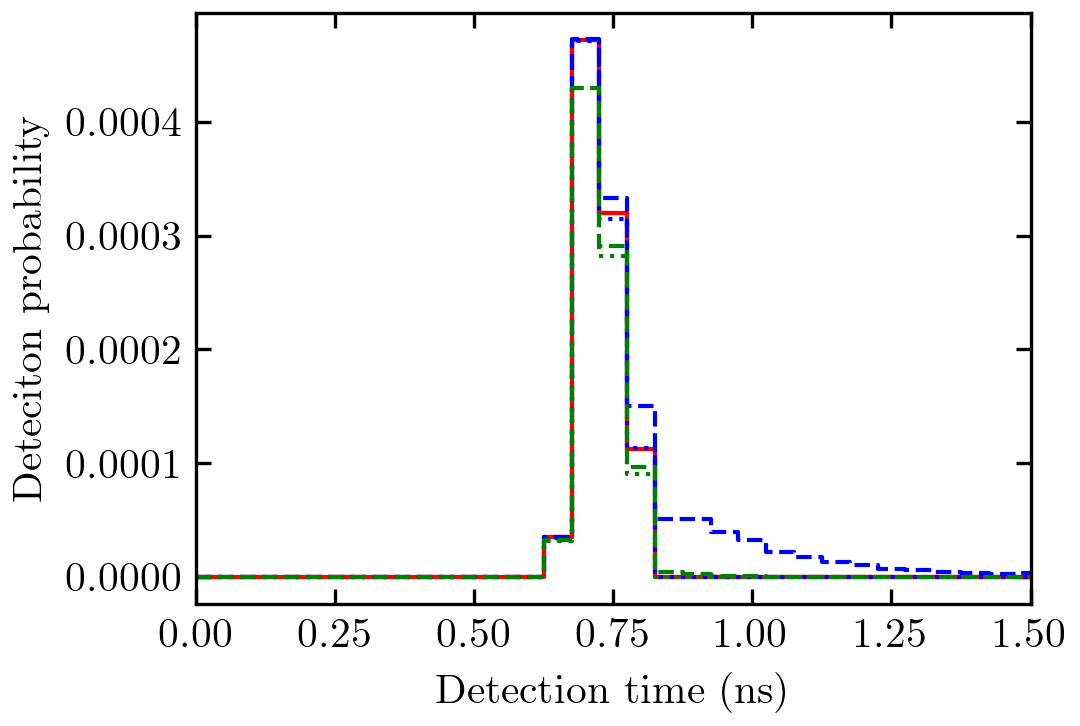}
        \caption{}
        \label{sfig:asm_vs_mc_0cm}
    \end{subfigure}
    \caption{System matrix elements without time resolution for detector position 73 and emission positions (a) ${z=\SI{-20}{cm}}$, and (b) ${z=\SI{0}{cm}}$. Red: \acs*{asm} with (solid line) and without (dash dot) attenuation of neighboring detectors; blue: \acs*{mcsm} simulated with only detector 73 (dots) or all 110 detectors (dash dash dot), and green: \acs*{mcsm} with \SI{1}{\mega\electronvolt} threshold on the detected energy).}
    \label{fig:asm_vs_mc_det73}
\end{figure}
\subsection{Detection efficiency}\label{ssec:det_efficiency}
When calculating the \ac{mcsm}, we applied no event selection unless stated otherwise. However, in simulations and experimental implementations of \ac{pgt} a threshold is often applied to $E_D$, the measured energy, to reduce the number of non-informative detection events. Hence, we used an energy threshold of \SI{1}{\mega\electronvolt} in some tests. To quantify the reduction in detection probability due to the energy threshold, we ran a simulation of a single detector facing a centered photon source. 
The ratio between the maximum detection probabilities of the \ac{mcsm} with and without the  \SI{1}{\mega\electronvolt} 
threshold was $0.91$. Consequently, we set $\mu_D/\mu=\num{0.91}$ in the \ac{asm} for comparisons with \acp{mcsm} calculated with $E_D>\SI{1}{\mega\electronvolt}$ and $\mu_D/\mu=\num{1}$ when no event selection was applied.

\subsection{Scatter and attenuation in multiple detectors}\label{ssec:intercrystal_s_and_a}
To study the effects of multiple closely-packed detectors, we simulated the detector geometry of \cite{ferrero2022frontiers} and \cite{Werner_2024}. A total of 110 detectors were arranged in five planes of 22 detectors each: nine along a straight line and thirteen in an arc. To isolate the impact of intercrystal scatter and attenuation in neighboring crystals, we simulated detector 73 (see \cref{sfig:maxerror_slice}) with and without the surrounding crystals present. In the \ac{asm} surrounding detectors were considered in \cref{eq:inter_crystalattenuation}. Intercrystal scatter was not modeled in the \ac{asm}. The results of the comparison are included in \cref{fig:asm_vs_mc_det73}. For a single detector, the \ac{asm} matches the \ac{mcsm} closely.
When including all neighboring crystals, a tail of late interactions appears in the \ac{mcsm}, while the \ac{asm} values drop to \SI{80}{\percent} of the \ac{mcsm} for emissions at ${z=\SI{-20}{\centi\meter}}$. The tail in the \ac{mcsm} is eliminated when imposing a \SI{1}{\mega\electronvolt} energy threshold. For all 110 detectors, the median absolute difference\footnote{Excluding errors of exactly 0, since the majority of system matrix elements is 0 when not considering a finite time resolution.} between \ac{asm} and \ac{mcsm} with $E_D>\SI{1}{\mega\electronvolt}$ is \SI{0.01}{\percent} and, as shown in \cref{fig:max_deviations_asm_vs_mcsm}, the difference does not exceed \SI{10.6}{\percent}. As demonstrated for detector position 73 in the inset of \cref{fig:max_deviations_asm_vs_mcsm}, the largest deviations correspond to underestimations of the \ac{asm} values where the paths between the emission point and the detector are substantially obstructed by another detector. The error for the line of detectors lies below \SI{3.5}{\percent}.

\subsection{Reconstructed distributions and stopping power}
\begin{figure}
    \centering
    \begin{subfigure}{.49\columnwidth}
        \centering
        \includegraphics[width=\linewidth]{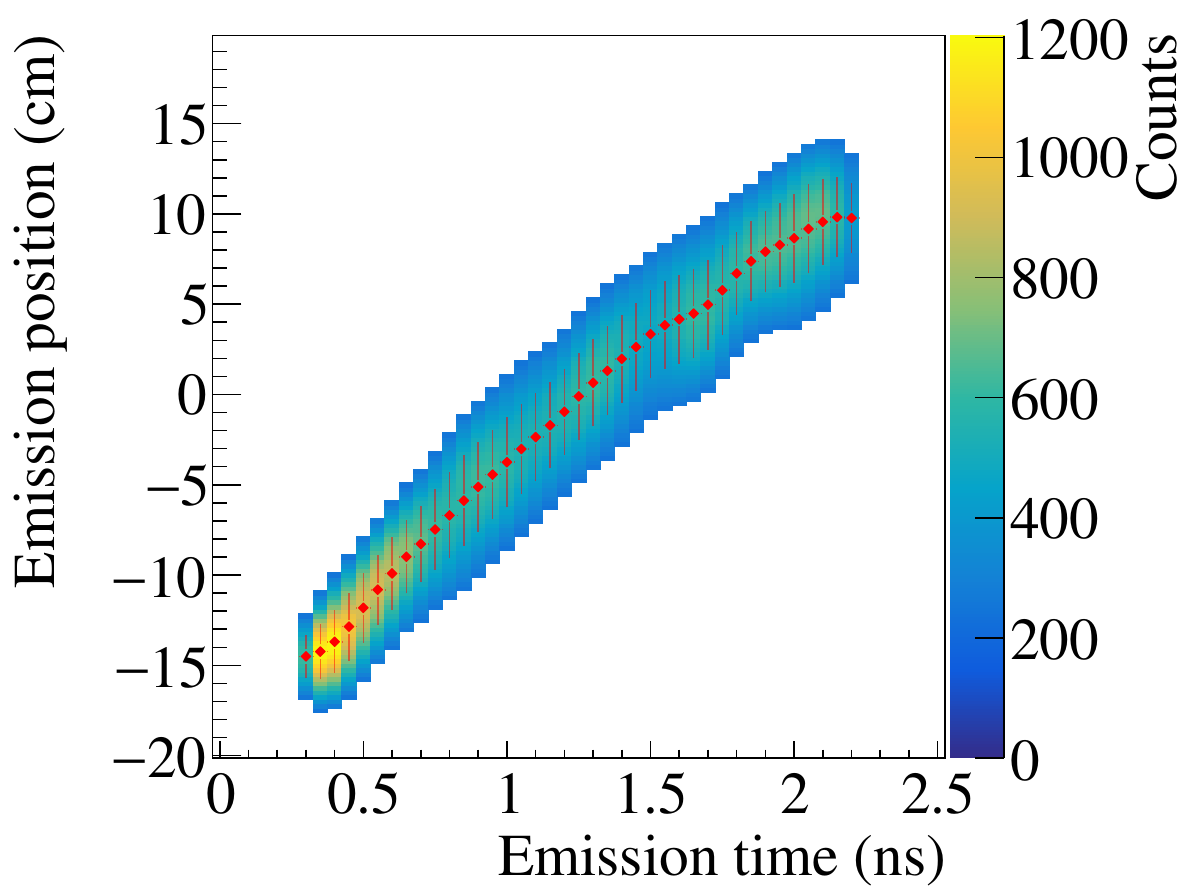}
        \caption{}
        \label{sfig:asm_219MeVreco}
    \end{subfigure}
    \begin{subfigure}{.49\columnwidth}
        \centering
        \includegraphics[width=\linewidth]{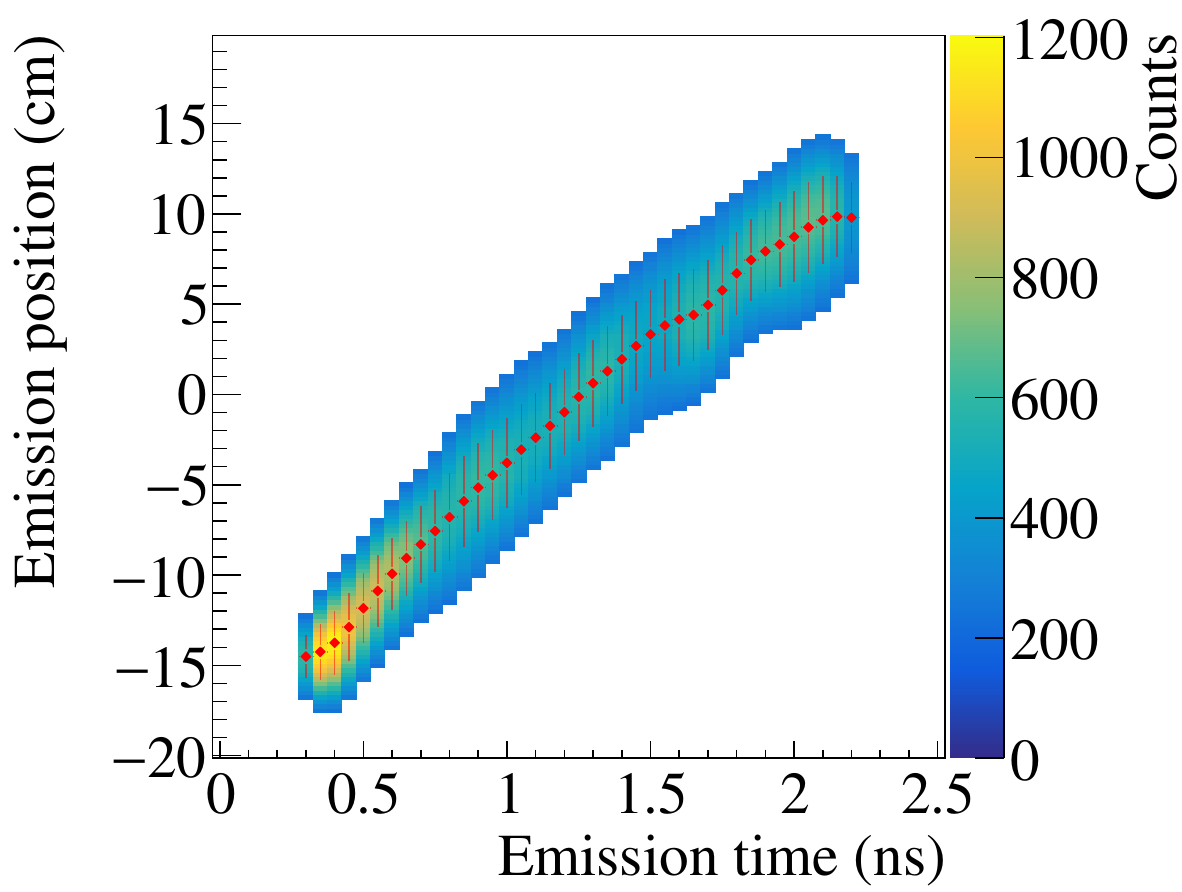}
        \caption{}
        \label{sfig:mcsm_219MeVreco}
    \end{subfigure}
    \begin{subfigure}{.49\columnwidth}
        \centering
        \includegraphics[width=\linewidth]{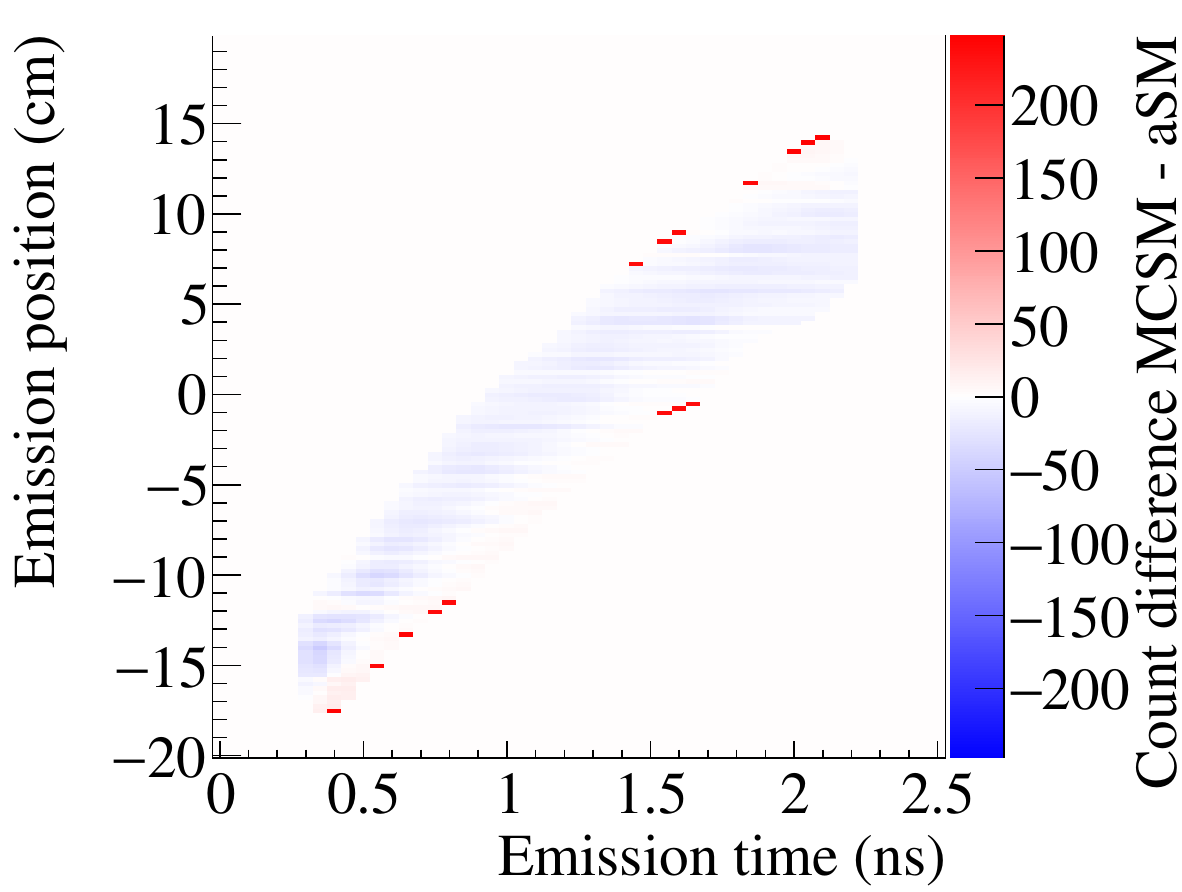}
        \caption{}
        \label{sfig:mcsm-asm_219MeVreco}
    \end{subfigure}
    \caption{Reconstruction of the spatiotemporal PG emission distribution of a simulated \SI{219}{\mega\electronvolt} proton beam using (a) \acs*{asm} or (b) \acs*{mcsm} after convolution with the \acs*{drf} (100 iterations, intensity threshold \SI{20}{\percent}). Red markers show the spatial \acs*{cog} over time and its standard deviation. (c) Difference between (a) and (b).}
    \label{fig:asm_vs_mc_recos}
\end{figure}
To evaluate the effect of the differences between the system models on the reconstructed distributions we used the Monte Carlo-simulated dataset from \cite{Werner_2024}. The beam particles were protons. The dataset contains 17 proton-beam energies impinging on a PMMA target (50 realizations per energy). To match the timestamp calculation of these simulations, we applied a convolution along the detection time with the \ac{drf} shown in \cref{sfig:timediff_MC} to the system models. To model the detector time resolution, a normal distribution with standard deviation \SI{106}{\pico\second} was added both to the data (event-by-event) and the system models (convolution). 
Almost no differences between the reconstructed distributions shown in \cref{fig:asm_vs_mc_recos} are visually discernible.
Quantitatively, the maximum reconstructed intensity increased slightly from 1180 (\acs*{mcsm}) to 1200 (\acs*{asm}). \Cref{sfig:mcsm-asm_219MeVreco} shows errors up to \num{240} counts, which are related to pixel intensities slightly above and below the intensity threshold. The intensity threshold applied to the distributions differs by 4 counts, since the threshold was defined relative to the maximum intensity of the given distribution. The maximum difference of non-zero intensities is 50 counts, which is \SI{4}{\percent} of the maximum reconstructed intensity.

To extract the stopping power from the reconstructed spatiotemporal distribution as in \cite{ferrero2022frontiers} and \cite{Werner_2024},
we first applied a threshold to the distribution and then calculated the spatial \acp{cog} of the \ac{pg} emission for each timestep (red markers in \cref{fig:asm_vs_mc_recos}). The free parameters of the primary particle motion model were obtained via non-linear regression:  
The model equations were fit to the position-time pairs. 
Next, the corresponding stopping power profile over depth within the phantom was calculated. As a range estimate, we used the \ac{cog} of the last emission bin (in time) above the threshold.
\begin{figure}
    \centering
    \begin{subfigure}{\columnwidth}
        \centering
        \includegraphics[width=\linewidth]{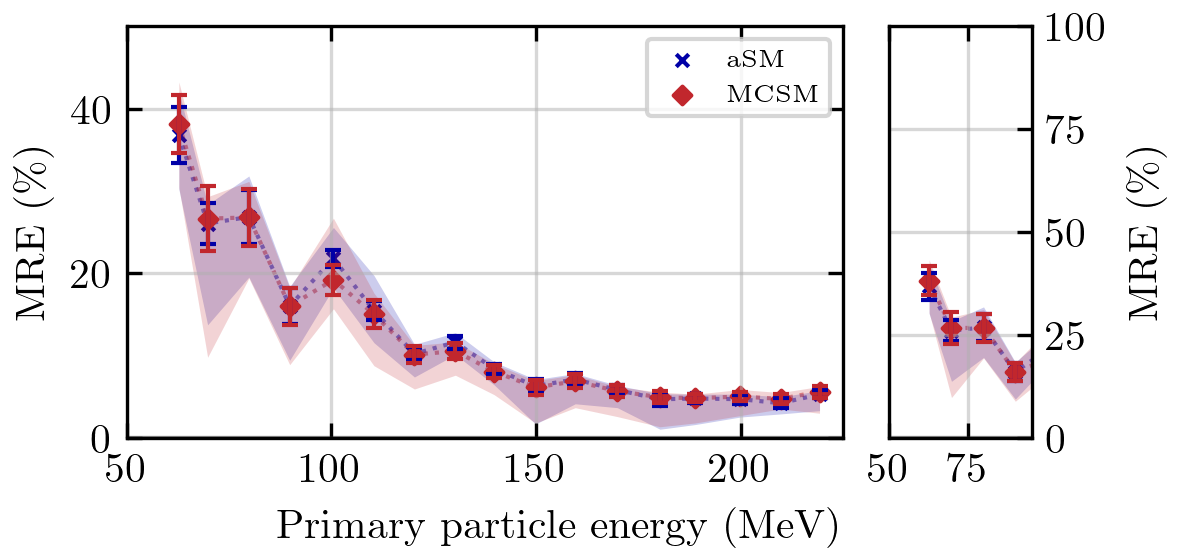}
        \caption{}
        \label{sfig:asm_vs_mc_mre}
    \end{subfigure}
    \begin{subfigure}{\columnwidth}
        \centering
        \includegraphics[width=\linewidth]{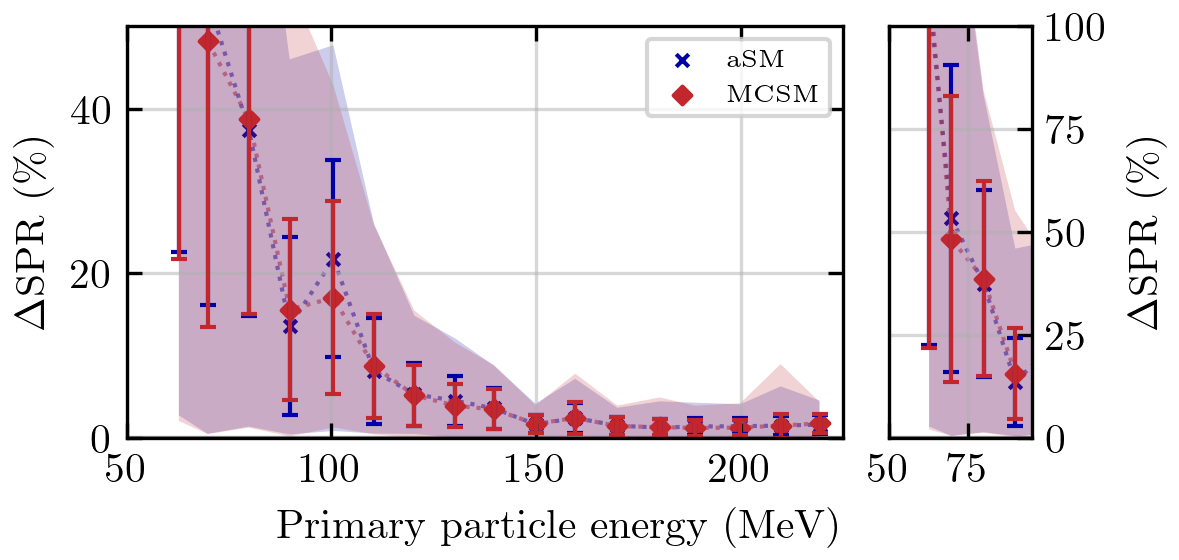}
        \caption{}
        \label{sfig:asm_vs_mc_spr}
    \end{subfigure}
    \caption{Average (marker), standard deviation (whiskers) and minimum and maximum (shaded area) of (a) \acs*{mre} and (b) \acs*{spr} error for reconstructions using \acs*{asm} and \acs*{mcsm} after convolution with the \acs*{drf}.}
    \label{fig:asm_vs_mc_mre_and_spr}
\end{figure}
The range and stopping power estimates are affected by the  number of iterations in the reconstruction as well as by the post-reconstruction threshold. To find the optimum values,
we applied the scenario-based approach optimizing the \ac{mre} introduced in \cite{Werner_2024}. The \ac{mre} compares the estimated stopping power to NIST PSTAR reference values for homogeneously spaced positions along the beam path. Additionally, we calculated the estimated \ac{spr} to water at \SI{100}{\mega\electronvolt}. These metrics showed that the stopping power estimations based on reconstruction with \ac{asm} and \ac{mcsm} were well within the statistical uncertainty (see \cref{fig:asm_vs_mc_mre_and_spr}). Compared to \cite{Werner_2024}, the standard deviation related to the \ac{mre} was reduced significantly. The \ac{spr} is slightly worse than reported in \cite{Werner_2024}, which is expected, since the optimization applied here only considered the \ac{mre}.

\section{Discussion}
We have shown that for single detectors the \ac{asm} is capable of reproducing the \ac{mcsm} values accurately. For many closely-packed detectors the error of some system matrix elements increased to \SI{10.6}{\percent}, but the median error was only \SI{0.01}{\percent}. As attenuation was included in the \ac{asm}, the remaining difference and the tail observed for the \ac{mcsm} in \cref{fig:asm_vs_mc_det73} (without energy threshold) can be attributed to intercrystal scatter. In the Monte-Carlo simulation, the detectors were placed in vacuum.
Therefore, any low-energy particle (e.g. electrons) leaving the detector where the first photon interaction happens had high chances of reaching another detector and contributing to the late-detection tail. The modeled attenuation in neighboring detectors based on $\mu$ was overestimated in the \ac{asm}, since intercrystal scatter offered an alternative route to detection. In any case, the late-detection tail is not relevant in experimental \ac{pgt} data, as the threshold is usually set above \SI{511}{\kilo\electronvolt} to avoid the contribution of positron-emitting isotopes, because they are not as closely correlated in time with the primary particles. The change in detection efficiency $\mu_D/\mu$ caused by applying an energy threshold was successfully included in the \ac{asm} based only on a \SI{30}{\second} simulation of a single emission position and detector.
Since intercrystal scatter is complex to model and highly sensitive to the relative placement of the detectors, adding intercrystal scatter to the \ac{asm} is beyond the scope of this work. 

Neither the \ac{mcsm} nor the \ac{asm} took scatter and attenuation within the phantom into account -- not only because of the computational complexity (especially for voxelized cases), but also because the underlying assumption of a given phantom or patient geometry may bias the results. One goal of treatment verification is to confirm the assumed patient anatomy. Therefore, robustness against possible deviations is key when including information about the phantom or patient geometry.
Using the dataset of \cite{Werner_2024}, we have demonstrated that the two models are equivalent in terms of reconstruction and stopping power estimation performance, i.e., the differences in some system matrix elements did not cause significant changes in the reconstructed distributions.

\subsection{Adjustments to experimental data}
The \ac{pg} energy spectrum consists of distinct peaks and a continuous background with the peak at \SI{4.4}{\mega\electronvolt} being one of the most prominent \cite{Verburg2012}. The full spectrum could be incorporated by calculating its effective $\mu$ and $\mu_D$ while keeping in mind, that the spectrum depends on the beam energy and tissue composition and might change throughout the detector as low-energy photons are prone to be attenuated faster than higher energy ones (so-called beam hardening). The implemented system models only considered photon energies of \SI{4.4}{\mega\electronvolt}.
Nevertheless, the successful results obtained from reconstructions based on the \ac{mcsm} applied to both simulated data with the full spectrum \cite{ferrero2022frontiers,Werner_2024}, as well as experimental data \cite{Ferrero2023IEEEexpReco,WERNER2024ECMP_expSP} 
shows this to be a reasonable simplification, as long as the attenuation coefficient is calculated using a sufficiently representative photon energy. 

Signal delay effects can be incorporated into the system models via a convolution with the \ac{drf} as demonstrated for the simulated \ac{pgt} spectra from \cite{Werner_2024}, although large shifts should be corrected at the data level since the observed detection time interval is limited. The \ac{drf} should also include the finite time resolution of the detector and effects of its readout. Depending on the detection system, several beam parameters need to be considered \cite{koegler2024modelPGT}.  Both system model implementations included \ac{pg} emissions only on the beam axis and did not model the emission time width $\Delta t_\point{P}$. Including the true $\Delta t_\point{P}$ as well as the transversal beam distribution $\Delta x$ and $\Delta y$ are possible improvements based on the theoretical framework summarized in \cref{eq:dP_P_tp_l_omega}. Also,  for the spatial emission bins other basis functions  instead of voxels could be implemented, since the \ac{asm} derivation is independent of the shape of the indicator functions up to \cref{eq:dP_P_tp_l_omega}, provided that the $L^1$ norm of the distributions is finite and non-zero.

The preliminary results with experimental data in \cite{Ferrero2023IEEEexpReco,WERNER2024ECMP_expSP} were achieved by incorporating the detector efficiency, a constant detector-dependent signal delay, and the coincidence time resolution into the \ac{mcsm}. Further details of the acquisition system were not needed to achieve reconstructions usable for stopping power estimation and the same adjustments could be applied to the \ac{asm}. Since the \ac{asm} and \ac{mcsm} showed good agreement element-wise as well as in reconstruction and stopping power estimation, we expect the \ac{asm} to also be applicable to experimental data.

\subsection{Use-cases}
The execution time for the \ac{asm} scales inversely with $\Delta\theta$, $\Delta\varphi$ as well as linearly with the number of detectors and the number of spatial emission and temporal detection bins. For the \ac{mcsm}, the execution time is mostly determined by the number of photons simulated, which directly affects the statistical error of the result. No significant errors caused by the numerical integration over emission directions in the \ac{asm} were observed. Depending on the detector geometry and positions, larger angle step sizes may be sufficient. More efficient sampling methods could be also considered. Even though the \ac{asm} does not lead to a closed solution for a system model, its fast execution time and minimal requirement for storage space make it useful for many studies related to \ac{pgt}-based reconstruction methods like \ac{serpgt} \cite{Pennazio_2022} and \ac{pgti} \cite{Jacquet_2021}. The \ac{asm} significantly speeds up investigations of new detector arrangements, optimization of detector configuration, as well as the calculation of a system model for off-center particle beams. Also, data-driven evaluations of \ac{pgt} data \cite{Werner_2019,schellhammer2022multivariate, KIESLICH2025machinelearning_temporal_spectral} may benefit from studying the effects of the detector placement, material and geometry using the \ac{asm} to predict the expected measurements. Finally, changing $\operatorname{\mathds{1}}_n$ to $\delta(t_D-t_n)$ in the \ac{asm} enables list-mode based \ac{serpgt}. \Acp{mcsm} can only estimate the detection probability for non-discretized measurements from counts in finite intervals. List-mode \ac{serpgt} would remove the need for discretizing the time measurement and, depending on the number of measured events, may speed up the reconstruction. In that case, $t_n$ would denote the detection time assigned to a given event, instead of the center of a detection time interval. 

\section{Conclusion}
We have derived and tested an \ac{asm} for \ac{serpgt}. It is approximately 1500 times more computationally efficient than the reference \ac{mcsm}. Although intercrystal scatter was not modeled in the \ac{asm}, no significant differences in performance were observed at the level of the \ac{serpgt} output and stopping power estimation.
While further improvements to the model are possible, the current version is ready to replace the \ac{mcsm} in upcoming studies of \ac{serpgt} and related techniques. The development of the \ac{asm} is a significant step towards the implementation and refinement of \ac{pgt}-based verification methods for clinically relevant scenarios.

\appendix
\label{sec:app_convolution}
The integration over $t_\point{P}$ to go from \cref{eq:dP_P_tp_l_omega} to $\prob_{\point{P},l,\Omega}$  can be expressed as a cross-corelation $\star$ of $\operatorname{\mathds{1}}_p$ and $\operatorname{\mathds{1}}_n$,
\begin{align}
    \int_{\mathbb{R}} \dd t_\point{P} &\operatorname{\mathds{1}}_{p}(t_\point{P})
    \operatorname{\mathds{1}}_{n}\left(\left(l+\diff{\point{Q}\point{P}}\right)/c+t_\point{P}\right) \nonumber\\
    &=\left( \operatorname{\mathds{1}}_{p} \star
    \operatorname{\mathds{1}}_{n}\right)\left(\left(l+\diff{\point{Q}\point{P}}\right)/c\right),
\end{align}
evaluated at time $\left(l+\diff{\point{Q}\point{P}}\right)/c$. A cross-correlation is equivalent to a convolution with the sign of one of the function arguments inverted. This leads to
\begin{align}
    \dd \prob_{\point{P},l,\Omega}&(d,n,j,p) \nonumber\\
    =&
    \dd V_\point{P} \frac{\operatorname{\mathds{1}}_j(\point{P})}{\Delta x \Delta y \Delta z}
    \frac{\dd\Omega}{4\pi}\,\dd l\, f_{d}(l) \operatorname{\mathds{1}}_{\point{P},\Omega,d}(l)\, \mu_D \nonumber\\&
    \frac{1}{\Delta t_\point{P}}\left( \operatorname{\mathds{1}}_{p} \star
    \operatorname{\mathds{1}}_{n}\right)\left(\left(l+\diff{\point{Q}\point{P}}\right)/c\right).
\end{align}
This formulation eases the prediction of the shape of the distribution of valid detection positions, $l$, given by the distribution of accepted emission and detection times, since it can easily be calculated based on known convolution results.

\balance

\bibliographystyle{IEEEtran}
\bibliography{literature}

\end{document}